\documentclass[prx,floatfix,showpacs,superscriptaddress,longbibliography,twocolumn]{revtex4-1}
\usepackage[USenglish]{babel}
\usepackage{amsmath,amssymb,amsfonts}
\usepackage{bm}
\usepackage{epsfig}
\usepackage{subfigure}

\usepackage[colorlinks=true,linkcolor=blue,citecolor=red]{hyperref}

\begin{document}

\date{\today}

\author{Antonio~Picano}
\affiliation{Department of Physics, University of Erlangen-N\"urnberg,
  91058 Erlangen, Germany}

\author{Martin~Eckstein}
\affiliation{Department of Physics, University of Erlangen-N\"urnberg,
  91058 Erlangen, Germany}

\title{Accelerated gap collapse in a Slater antiferromagnet}

\begin{abstract}
We study the melting of long-range antiferromagnetic order in the Hubbard model after an interaction quench, using non-equilibrium dynamical mean-field theory. From previous studies, the system is known to quickly relax into a prethermal symmetry-broken state. Using a convergent truncation of the memory integrals in the Kadanoff Baym equations, we unravel the subsequent relaxation dynamics of this state over several orders of magnitude in time. At long times, the prethermal state can be characterized by a single slow variable which is related to the conduction band population. The dynamics of this variable does not follow the paradigmatic steady relaxation of pre-thermal states: It is highly nonlinear, with a pronounced speedup once the gap falls below a certain value. This behavior indicates that non-thermal order can be self-stabilized on some timescale. It is not reproduced using simple Fermi's golden rule estimate for the evolution of the conduction band population.
\end{abstract}
\pacs{}
\maketitle

{\em Introduction --} 
The ultrafast quantum dynamics of symmetry-broken states is a problem at the foundations of quantum many-particle physics, which has become accessible in experiments with cold atoms \cite{GuardadoSanchez2018} as well as through the photo-induced dynamics of condensed matter phases, including charge density waves \cite{Schmitt2008, Schaefer2010, Rohwer2011, Hellmann2012, Rettig2016}, spin density waves \cite{Kim2012, Rettig2012, Naseska2018}, or excitonic insulators \cite{Mathias2016, Mor2017}. While many experiments in the solid use pulses that are strong enough to quench the order almost instantly, and focus on the recovery,  several examples  have shown that long-range order can be transiently enhanced even if the excitation energy is sufficient to melt the order after thermalization  \cite{Kim2012,Mor2017, Rettig2016}. In other cases, the excitation can also initiate transitions to entirely new long-lived phases \cite{Beaud2014, Stojchevska2014, Fausti2011a}.  A generic mechanism to generate such non-thermal states is unknown,  but one intriguing possibility is that the thermalization bottleneck may be determined by the  insulating gap resulting from the long-range order itself, thus opening the room for highly non-linear dynamics and self-sustained non-thermal phases.

Non-thermal symmetry broken states can be an example of prethermalization  \cite{Berges2004, Moeckel2008, Eckstein2009, Werner2012,Barmettler2009, Polkovnikov2011, Babadi2015, Tsuji2013, Langen2016}, which more generally refers to a quick relaxation to a non-thermal state followed by thermalization on much longer timescales. The collapse of long-range order after an excitation is therefore closely related to the question of how prethermal states thermalize. One can view prethermal states as having relaxed under constraints imposed by approximate conservation laws \cite{Kollar2011, DAlessio2016, Langen2016}, and describe them in terms of a  statistical ensemble with corresponding generalized chemical potentials \cite{Jaynes1957, Rigol2006}. These almost conserved quantities  provide {\em slow variables} in the evolution, whose relaxation can in many cases be  understood from estimates based on Fermi's golden rule \cite{Moeckel2008, Stark2013, DAlessio2016, Mallayya2019}. In insulators with a robust gap, such as Mott insulators, this paradigm seems to apply: There is a clear separation of the time to thermalize an excited distribution function {\em within} the energy bands, and the redistribution of spectral weight between the bands through Auger scattering. As a result, the individual band populations provide the slow variables, and the prethermal state can be described in terms of a  universal Fermi distribution with separate chemical potential in the occupied and unoccupied bands \cite{Li2020, He2015}. If the gap is due to long-range order, the situation is much less clear. The gap will close with the melting of the order, giving rise to a potentially highly nonlinear dynamics. Moreover, the relaxation dynamics can exhibit non-thermal critical behavior separating regimes of ordered and disordered prethermal states \cite{Langen2016,Berges2008, Nowak2014,Tsuji2013}, and it is unclear in general whether also the thermalization slows down in such a critical region.  

In the present work we use non-equilibrium dynamical mean-field theory (DMFT) \cite{Aoki2014} to study the interaction quench in the antiferromagnetic Hubbard model as a paradigmatic example for the excited state dynamics in a symmetry-broken phase at weak-coupling. The prethermalization in this model has been shown to exhibit regions of non-thermal criticality \cite{Tsuji2013}. Here we mainly focus on a regime in which the prethermal state is still ordered,  and aim to understand its nature as well as the subsequent thermalization and gap closing. 
This study requires to extend the simulations of the quantum dynamics almost two orders of magnitude longer times than in related previous studies \cite{Tsuji2013}. This numerical challenge, which rests on the large separation of timescales inherent in the problem, is overcome by a controlled truncation of the memory time in the Kadanoff-Baym equations \cite{Schueler2018}. 

{\em Model --}
We study the half-filled Hubbard model
\begin{align}
\label{ham:hubb}
H = -t_h \sum_{\langle i,j \rangle,\sigma} c_{i\sigma}^\dagger c_{j\sigma} + U(t)\sum_{j} (n_{j\uparrow}-\tfrac12)(n_{j\downarrow}-\tfrac12)
\end{align}
 on a bipartite lattice. Here $c_{j\sigma}$ ($c_{j\sigma}^\dagger$) denotes the annihilation (creation) operator for a Fermion with spin $\sigma\in\{\uparrow,\downarrow\}$ on lattice site $j$. The first term in Eq.~\eqref{ham:hubb} describes hopping with amplitude $t_h$ between nearest neighbor sites $\langle i,j\rangle$, the second term, with $n_{j\sigma}=c_{j\sigma}^\dagger c_{j\sigma}$, is the local interaction. We evaluate in particular the Ne\'el order parameter  $m(t) = \frac{1}{2}( n_{A,\uparrow}-n_{B,\uparrow})= \frac{1}{2}( n_{B,\downarrow}-n_{A,\downarrow})$ where $n_{\alpha,\sigma}$ is the density of spin $\sigma$ on a site $j$ within the sublattice $\alpha\in\{A,B\}$ of the bipartite lattice. The actual simulations assume a semi-elliptic local density of states $D_0(\epsilon)=\sqrt{4-\epsilon^2}/(2\pi)$ for the noninteracting model, corresponding to a Bethe lattice with hopping $t_h=1$. The unit of energy is $1/4$ of the bandwidth, and its inverse defines the unit of time ($\hbar=1$).

The DMFT equations are the same as in Ref.~\cite{Tsuji2013}, and are therefore only reproduced in the appendix. They are solved using the NESSi simulation package \cite{Schueler2020}, using second order self-consistent perturbation theory as an impurity solver \cite{Tsuji2013a}.  The main numerical challenge in long-time simulations are the memory integrals in the Kadanoff-Baym equations \cite{Aoki2014}. However, because the double-time correlation self-energy $\Sigma(t,t')$ decays quickly as a function of relative time, one can set $\Sigma(t,t')=0$ for $|t-t'|>t_c$, where the cutoff time $t_c$ is varied until the results do no longer depend on $t_c$.  In the present case, a cutoff $t_c=20$ is sufficient to have converged results at all times, up to $t\approx10^4$. Further numerical details are discussed in the appendix.

\begin{figure}[tbp]
\centerline{\includegraphics[width=0.5\textwidth]{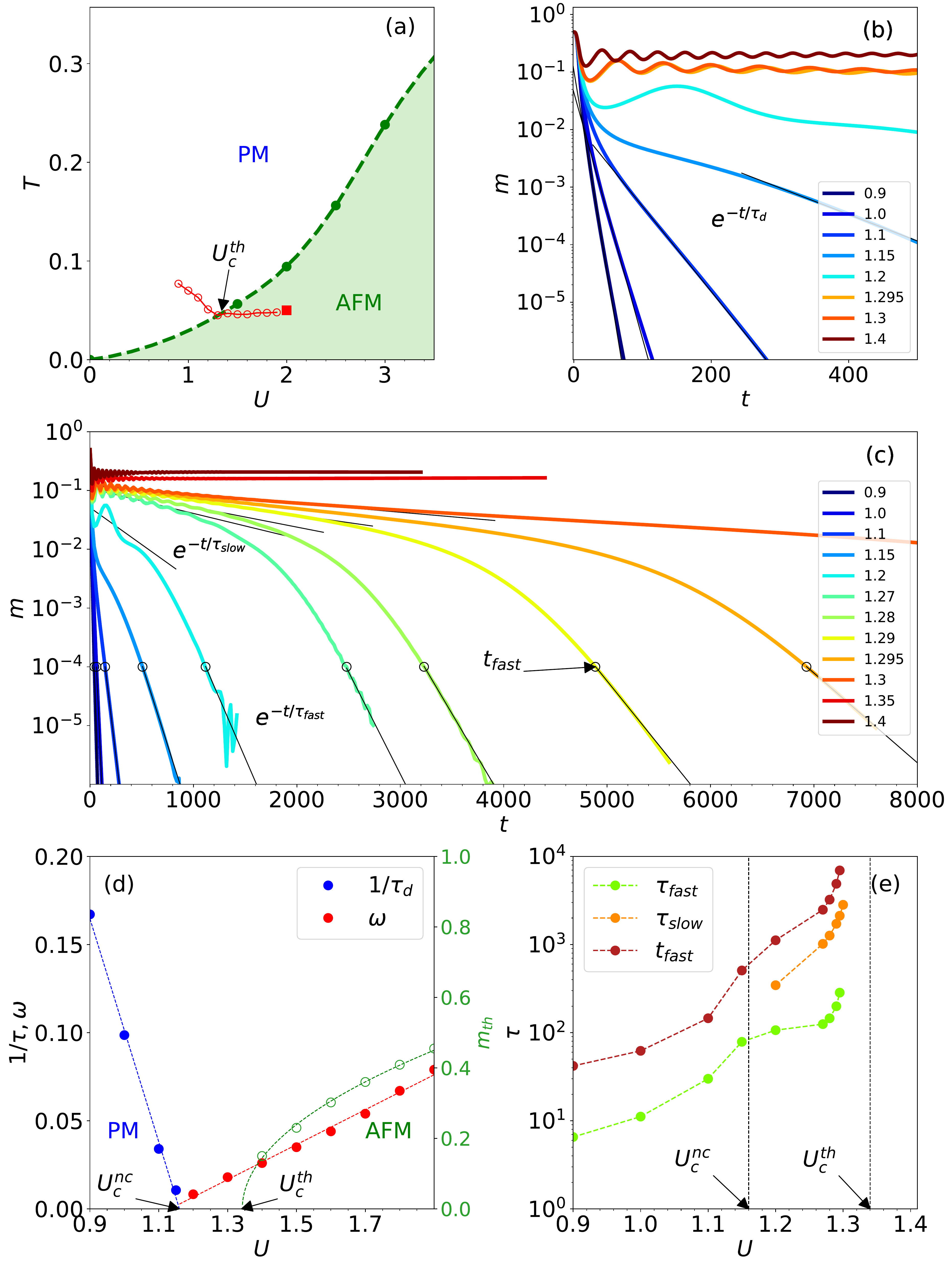}}
\caption{
a) Equilibrium phase diagram. The filled square denotes the initial state at $U=2$, open circles show the temperature $T_{\rm eff}$ after a quench and possible thermalization. b) and c) Order parameter $m(t)$ after quenches to different values of $U$. d) Timescales to characterize the evolution in b) and c), extracted from the fits as indicated by various dashed lines and described in the main text. Open circles show the order $m_{th}$ after thermalization.}
\label{fig01}
\end{figure}

{\em Results --}
Figure~\ref{fig01}a shows the phase diagram of the model, with the Ne\'el phase ($|m|>0$) below $T_N$. For time $t<0$, we choose a symmetry-broken thermal equilibrium state at interaction $U_i=2$ and temperature $T=1/20<T_N$, see red square. At $t>0$, the interaction is quenched to a different value $U<2$. The energy is conserved after the quench, and the open symbols in Fig.~\ref{fig01}a show for each $U$ the temperature $T_{\rm eff}$ of an equilibrium system with the same energy density, i.e., the temperature after a possible thermalization. In Fig.~\ref{fig01}d we show the value $m_{th}$ in the corresponding equilibrium state at $T=T_{\rm eff}$. One can see that there is a threshold $U_c^{th}\approx1.3$, so that one would expect thermalization to a disordered state for large quenches ($U<U_c^{th}$), and to an ordered state for smaller quenches ($U>U_c^{th}$).   In the following we mainly focus on quenches $U<U_c^{th}$. Figure~\ref{fig01}b and c show the dynamics of $m$ for various values of $U$. For  $U\lesssim 1.15$ we observe a rather quick exponential decay of $m(t)$, so that the relevant dynamics is over after few $100$ inverse hoppings (Fig.~\ref{fig01}b). For $U\gtrsim 1.15$, in contrast, the initial dynamics shows a quick drop to a nonzero value of $m$, together with damped oscillations. This crossover around $U_{c}^{nc}=1.15$ corresponds to the non-thermal critical region described in \cite{Tsuji2013}, separating ordered and disordered prethermal states.  In agreement with Ref.~\cite{Tsuji2013}, both the decay rate $1/\tau_{d}$ extracted from an exponential fit $m(t)=Ae^{-t/\tau_{d}}$ (dashed lines in Fig.~\ref{fig01}b), and the frequency $\omega$ of the oscillations vanishes like $|U-U_{c}^{nc}|$ as the transition is approached (see Fig.~\ref{fig01}d).

For quenches $U>U_{c}^{nc}$, the system remains in the ordered phase for a long time, with a slow exponential decay of $m(t)$, until the system undergoes a rather sudden speedup of the decay (Fig.~\ref{fig01}c). To characterize this nonlinear dynamics empirically, we extract three timescales: A fit $m(t)= Ae^{-t/\tau_{\rm slow}}$ in the range where $m>0.05$ measures the slow decay time $\tau_{\rm slow}$ in the prethermal state; and a fit $m(t)=m_0 e^{-(t-t_{\rm fast})/\tau_{\rm fast}}$ to the data at $m<m_0$, with $m_0=10^{-4}$, measures the late thermalization time $\tau_{\rm fast}$ and the onset $t_{\rm fast}$ of the fast dynamics. All timescales show a slowdown as $U$ approaches the thermal threshold $U_{c}^{th}$ (Fig.~\ref{fig01}e), but  the trapping time $t_{\rm fast}$ is more than an order of magnitude longer than the later fast thermalization $\tau_{\rm fast}$. This seemingly self-sustained trapping in the nonthermal state, with a later speedup of the dynamics, is the main numerical finding of this work, which will be characterized below. We emphasize that this is entirely different from the relaxation of other prethermal states, which are often found to evolve steadily towards thermal equilibrium \cite{Moeckel2008, Stark2013, DAlessio2016, Mallayya2019}. 

As a side remark, we can comment on the interesting question whether in the system at hand the vicinity of a nonthermal critical point would delay the subsequent thermalization. One can see that the onset $t_{\rm fast}$ of the fast thermalization does indeed show an anomaly in the nonthermal critical region $U\approx U_c^{nc}$, but it monotonously increases throughout the whole range. Therefore, in the present setting this onset of thermalization provides a natural cutoff for the critical behavior $\tau_{d}\sim|U-U_{c}^{nc}|^{-1}$, indicating no true criticality but a pass-by of a nearby critical point in parameter space, e.g., the critical point in the mean-field model \cite{Tsuji2013}.

\begin{figure}[tbp]
\centerline{\includegraphics[width=0.5\textwidth]{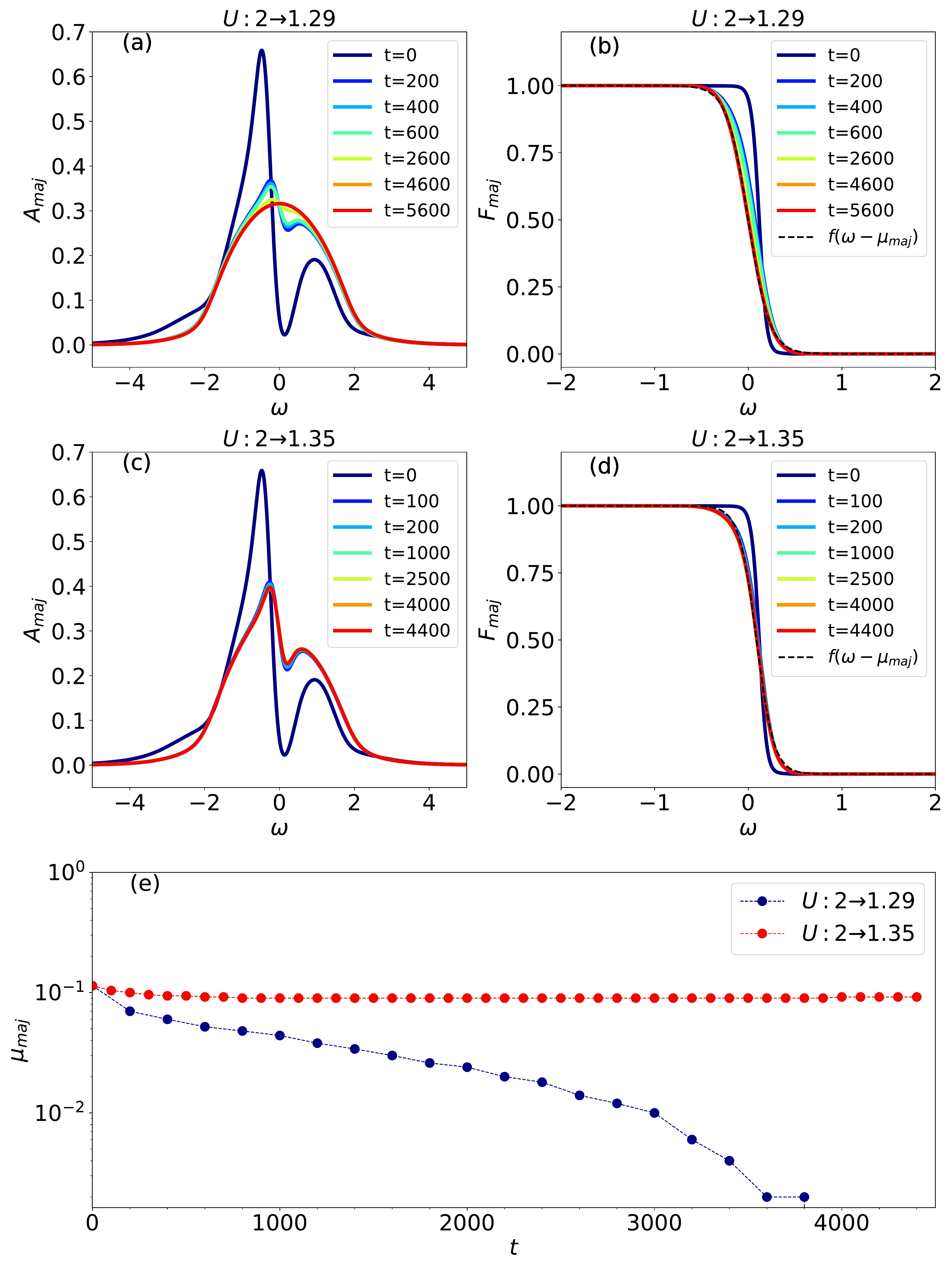}}
\caption{
Majority spin spectral function $A_{\rm maj}(\omega,t)$ (a) and (c) and occupation function 
$F_{\rm maj}(\omega,t)$ (b) and (d) for various times, as indicated by the color. The black dashed line is the final equilibrium state, the blue line the initial state.
e) Chemical potential $\mu_{\rm maj}$ extracted from a fit of $F_{\rm maj}(\omega,t)$ to a Fermi distribution. 
}
\label{fig02}
\end{figure}

To further characterize the prethermal state, we analyze the spin and site-dependent local spectral function
\begin{align}
A_{j\sigma}(\omega,t)
=
-\frac{1}{\pi}\text{Im}\int
ds \,e^{i\omega s } G_{j\sigma}^R(t+s/2,t-s/2).
\label{spectre}
\end{align}
An analogous expression for the  occupied spectrum $N_{j\sigma}(\omega,t)$, and the distribution function $F_{j\sigma}(\omega,t)=N_{j\sigma}(\omega,t)/A_{j\sigma}(\omega,t)$ is obtained by replacing the retarded Green's function $G^R$ with the lesser component in this expression. For the Ne\'el symmetry, we denote the spectrum for the majority spin ($\uparrow$ for site $j$ on the $A$ sublattice, or $\downarrow$ for $j$ on $B$)  by $A_{\rm maj}$, and  the minority spin by $A_{\rm min}$.  For technical reasons there is a  finite cutoff $s\in(-t_c,t_c)$, $t_c=20$, in Eq.~\eqref{spectre}, limiting the frequency resolution. Figure~\ref{fig02}  shows the majority spectrum and occupation function exemplarily for two quenches to $U=1.29$ ($U>U_c^{th}$), where the gap remains open, and to $U=1.35$ ($U_c^{nc}<U<U_c^{th}$), where a closing of the gap is observed.   The distribution function can be  described well with a Fermi function $F_{\rm maj}(\omega)=1/(e^{(\omega-\mu_{\rm maj})/T_{\rm maj}}+1)\equiv f(\omega-\mu_{\rm maj},T_{\rm maj})$ and a chemical potential $\mu_{\rm maj}>0$. The minority spin spectrum is symmetric, $A_{\rm min}(\omega)=A_{\rm maj}(-\omega)$, and therefore has a distribution function $F_{\rm min}(\omega)$ with $\mu_{\rm min}=-\mu_{\rm maj}$ and $T_{\rm min}=T_{\rm maj}$. Hence the full state can be described as a state with two separate chemical potentials, which relax towards equilibrium $\mu_{\rm min}=\mu_{\rm maj}=0$ with the decay of the antiferromagnetic order (Fig.~\ref{fig02}e). 

\begin{figure}[tbp]
\centerline{\includegraphics[width=0.49\textwidth]{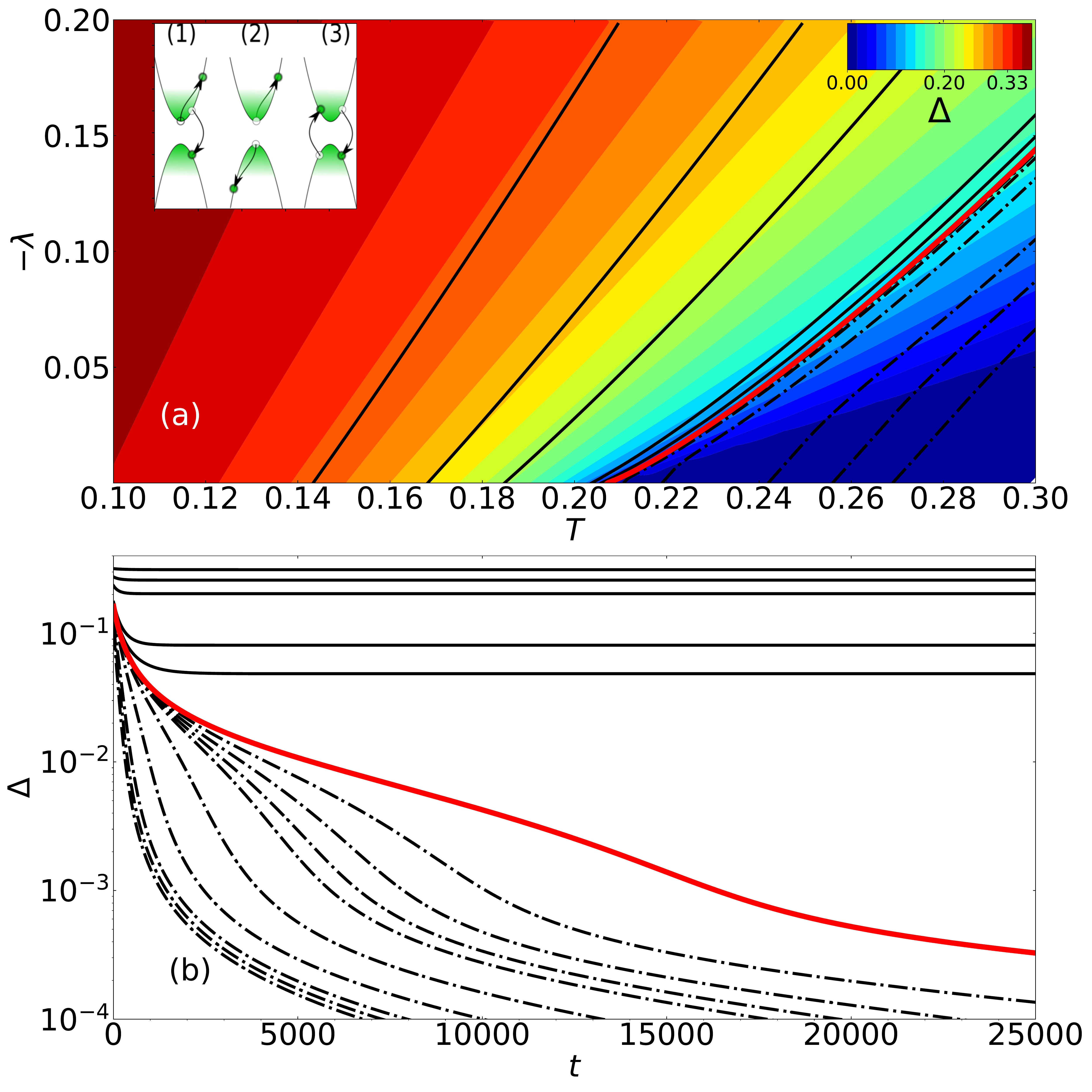}}
\caption{a) Non-equilibrium mean-field phase diagram: Gap $\Delta(T,\lambda)$ obtained from the gap equation \eqref{mindthegap}. The black lines correspond to lines of constant energy. The inset illustrates Auger (1) and intra-band (2),(3) scattering processes. b) Evolution of the prethermal state along the same constant energy lines as drawn in a). Lines from left to right in a) correspond to lines from top to bottom in b). The red lines in a) and b) indicate the evolution closest to the thermal threshold.
}
\label{fig03}
\end{figure}

In general, the prethermal nature of a state can be understood as a consequence of nearly conserved quantities, which impose constraints on the dynamics. The universal form of the distribution function is a numerical evidence that on the relevant timescales the prethermal state is determined largely by one constraint in addition to the energy conservation, which is the slow redistribution of the occupation across the gap. A starting point for the understanding of the prethermal state in the present example, in particular how two separate chemical potentials in the upper and lower band  lead to an {\em increased} gap with respect to an equilibrium state with the same total energy, is therefore provided by a mean-field solution, in which such slow variables are {\em exactly} conserved. The standard mean-field solution of the Slater antiferromagnet yields the equation for the gap $\Delta=U|m|$ 
\begin{align}
\frac{1}{U}
&=
\int_0^{\infty} d\epsilon D_0(\epsilon) 
\frac{F(-E)-F(E)}{E},
\label{mindthegap}
\end{align}
where $E=\sqrt{\epsilon^2+\Delta^2}$ is the mean-field dispersion, and $F(E)$ is the occupation function (for a derivation, see the appendix). In equilibrium, $F(E)$ is the Fermi distribution $f(E,T)$, but the validity of the gap equation is not restricted to equilibrium states. Figure~\ref{fig03}a shows the solution $\Delta(\lambda,T)$ for a non-thermal distribution with  chemical potentials $\mu_+=\lambda$ and and $\mu_-=-\lambda$ in the upper and lower band respectively, i.e., $F(E)=f(E-\lambda,T)$ for $E>0$ and $F(E)=f(E+\lambda,T)$ for $E<0$.  Solid and dashed lines in Fig~\ref{fig03}a indicate contour lines of constant total energy, $\lambda=0$ is the equilibrium state. For $\lambda<0$, $\Delta$ increases as one moves away from equilibrium along a constant energy contour, and Eq.~\eqref{mindthegap} has a nonzero solution irrespective of temperature. Mathematically, this is because for $\lambda<0$ the integrand in \eqref{mindthegap} has a logarithmic divergence as $\Delta\to 0$. Physically, for $\lambda<0$, the distribution is pushed away from the band bottom, where it is most harmful for the order. An analogous redistribution of the occupation explains the enhancement of $T_c$ in a superconductor using  microwave radiation \cite{Eliashberg1970}.

Assuming a separate chemical potential $\lambda$ for the upper and lower band corresponds to taking  the occupation $N_+=\int_0^\infty d\epsilon D_0(\epsilon) F(E)$ in the upper band as a conserved quantity, in addition to the energy and particle number. One may therefore ask to what extent the nonlinear {\em dynamics} observed in Fig.~\ref{fig01}c can be captured by allowing for scattering processes that lead to a slow evolution of $N_+$. In the spirit  of a slow evolution of almost conserved quantities \cite{Mallayya2019}, the evolution of $T(t)$, $\lambda(t)$ and $\Delta(t)$ should  be described by the self-consistent solution of four equations, which are (i) the conservation of the total energy $\mathcal{E}(\lambda(t),T(t))=\text{const.}$, (ii) the gap equation \eqref{mindthegap} which sets $\Delta(T,\lambda)$, (iii), the relation $N_+(\lambda,T)$, and (iv), an estimate for the rate  $dN_+/dt$ obtained from Fermi's Golden rule, 
\begin{align}
&\frac{dN_+}{dt}
=
\int d1d2d3d4 \, \Delta_{1234} \, |M_{1,2,3,4}|^2  \,\,\,\times
\nonumber
\\
&\times\,\,\,(1-F_{1})  F_{2} (1-F_{3})  F_{4}\,
\delta (E_{1}-E_{2}+E_{3}-E_{4}).
\label{rate}
\end{align}
The latter equation sums over all scattering processes from occupied orbitals $2$, $4$ to unoccupied orbitals $1$, $3$. $j=(\epsilon_j,\gamma_j)$ is short for the mean-field orbitals with energy $E_j=\gamma_j(\epsilon_j^2+\Delta^2)^{1/2}$  in the upper ($\gamma=+$) and lower ($\gamma=-$) band, $\int dj = \sum_{\gamma_j=\pm}\int d\epsilon_j D_0(\epsilon_j)$ is the sum over all states. The rate is determined by the occupation factors $F_j=F(E_j)$, the matrix element $M$ of the local interaction in the mean-field orbitals (see appendix), and a factor $\Delta _{1234}$ which measures by how much a scattering process changes $N_+$: Of relevance are Auger processes, i.e., process (1) in the inset of Fig.~\ref{fig03}, but not processes (2) and (3) which contribute only to intra-band thermalization. In Fig.~\ref{fig03}b we have solved the Boltzmann kinetics for different initial conditions. The system evolves along the constant energy lines in the $(T,\lambda)$ phase diagram either to an ordered or disordered equilibrium state (see solid and dashed lines in Fig.~\ref{fig03}a), respectively. The order of magnitude for the decay is in fact comparable to the slow times $\tau_{\rm slow}$ in the full dynamics, with a slowdown at the thermal threshold. On the other hand, the accelerated gap collapse is not captured by the Boltzmann analysis. 
This clearly demonstrates that correlation effects beyond mean-field in the electronic spectra are needed to obtain the long-time dynamics correctly even at a qualitative level. In the present model, such correlation effects become manifest in a filling-in of the gap in addition to the gap narrowing due to the reduction of the order parameter.


{\em Conclusion --}
In conclusion, we analyzed the slow melting of prethermal symmetry broken states after an interacting quench in the Hubbard model. We find that the distribution function in these states takes a universal form which is characterized by a slowly evolving chemical potential and temperature, indicating a single relaxation constraint. Moreover, we find a highly non-linear gap closing dynamics, in which a slow evolution of the order parameter $m(t)$ quickly transits into a faster gap collapse after the gap has fallen below a threshold. While the initial slow evolution is captured qualitatively by a mean field theory for the prethermal state with slowly evolving parameters through scattering, 
the rapid gap collapse is beyond this reasoning. 
Beyond providing a scenario for the thermalization of prethermal states which differs from a more conventional steady relaxation,
these results should be of interest to quench experiments in cold atoms \cite{GuardadoSanchez2018}. In condensed matter, they provide an intriguing hint at a relatively long stability of light-induced phases which are based on distributional engineering \cite{Mor2017}. Such long-lived states may also be a first stage to 
stabilize truly metastable nonequilibrium phases
\cite{Beaud2014, Stojchevska2014}. Most fascinating would be a direct observation of the accelerated gap collapse itself in real time. Indirect observation based on the fluence dependence of the gap-closing time has already lead to the conclusion that impact-ionization processes can imply an accelerated gap closure \cite{Mathias2016}.

We thank Philipp Werner for valuable discussions. This work was supported by the ERC starting grant No. 716648. The authors gratefully acknowledge the compute resources and support provided by the Erlangen Regional Computing Center (RRZE).

\appendix


\section{Mean-field equations}
\label{Sec:MF}

\subsection{Gap equation}

Within a suitable mean-field decoupling, the interaction term in Eq.~\eqref{ham:hubb} is replaced by (up to constants) 
\begin{align}
&(n_{j\uparrow}-\tfrac12)(n_{j\downarrow}-\tfrac12)
\to
n_{j\uparrow} \langle n_{j\downarrow}-\tfrac12\rangle
+
n_{j\downarrow} \langle n_{j\uparrow}-\tfrac12\rangle.
\end{align}
With the Ne\'el order parameter 
\begin{align}
\label{neel}
m
&=
\frac{1}{2}\langle n_{A,\uparrow}-n_{B,\uparrow}\rangle
=
\frac{1}{2}\langle n_{B,\downarrow}-n_{A,\downarrow}\rangle
\\
\label{mmh2j2j2}
&=\frac{\sigma}{2}\langle n_{A,\sigma}-n_{B,\sigma}\rangle
\end{align}
we have
\begin{align}
n_{j\uparrow} \langle n_{j\downarrow}-\tfrac12\rangle
+
n_{j\downarrow} \langle n_{j\uparrow}-\tfrac12\rangle
&=
-m \sum_{\sigma}
n_{j\sigma}\, \sigma  \xi_j,
\end{align}
where $\xi_j=+1(-1)$ for $j$ on the  $A(B)$ sublattice. For the momentum representation we use the magnetic Brillouin zone (MBZ) with two atoms per unit cell,
\begin{align}
c_{j,\sigma} &= \frac{1}{\sqrt{L}}\sum_{{\bm k}} c_{{\bm k},\alpha,\sigma} e^{i{\bm k}\vec{r_j}},
\end{align}
where  $\alpha\in\{A,B\}$ denotes the sublattice of site $j$, and $L$ is the number of momentum points in the MBZ. 
After introducing the spinors 
\begin{align}
\label{spinors}
\Psi_{{\bm k},\uparrow}
&=\begin{pmatrix} c_{{\bm k},A,\uparrow}\\ c_{{\bm k},B,\uparrow}\end{pmatrix},
\,\,\,\,
\Psi_{{\bm k},\downarrow}
=\begin{pmatrix} c_{{\bm k},B,\downarrow}\\ c_{{\bm k},A,\downarrow}\end{pmatrix},
\end{align}
the mean-field Hamiltonian is given by
\begin{align}
\label{ham:mf}
H_{mf}
&=\sum_{{\bm k},\sigma} 
\Psi_{{\bm k},\sigma}^\dagger h_{{\bm k}} \Psi_{{\bm k},\sigma},
\,\,\,\,h_{{\bm k}}
=\begin{pmatrix} -\Delta & \epsilon_{{\bm k}}\\\epsilon_{{\bm k}}& \Delta\end{pmatrix},
\end{align}
where $\epsilon_{{\bm k}}$ is the band dispersion of the Hamiltonian \eqref{ham:hubb}, and $\Delta=Um$.
Note that we have chosen an opposite order of the sublattice ($A,B$) components in the definition of the spinors $\Psi_{{\bm k},\downarrow}$ and  $\Psi_{{\bm k},\uparrow}$ in Eq.~\eqref{spinors}, so that the matrix $h_{{\bm k}}$ does not depend on spin.  

Because of particle-hole symmetry on a bipartite lattice, the dispersion is symmetric in energy, and one can always choose the MBZ such that it contains the positive (or negative) band energies, $MBZ=\{{\bm k}\in\text{full BZ}: \epsilon_{{\bm k}}\ge0\}$. Hence momentum summations are rewritten as  energy integrals
\begin{align}
\label{dos}
\frac{1}{L}\sum_{{\bm k}\in MBZ} g(\epsilon_{{\bm k}})
\to
2\int_0^\infty d\epsilon D_0(\epsilon) g(\epsilon).
\end{align}
Here $D_0(\epsilon)$ is the density of states of the original lattice, which is normalized to $\int d\epsilon D_0(\epsilon)=1$, so that the factor $2$ on the right-hand side of Eq.~\eqref{dos} ensures that both sides of the equation give one for $g(\epsilon)=1$. In the main text, the density of states  is assumed to have a semi-elliptic form $D_0(\epsilon)=\sqrt{4-\epsilon^2}/(2\pi)$. 

In terms of the spinors \eqref{spinors}, the order parameter \eqref{neel} is given by 
\begin{align}
m
&=
\frac{1}{2L}\sum_{{\bm k}} \Psi_{{\bm k},\uparrow}^\dagger \tau_{z} \Psi_{{\bm k},\uparrow}
=
\frac{1}{2L}\sum_{{\bm k}} \Psi_{{\bm k},\downarrow}^\dagger \tau_{z} \Psi_{{\bm k},\downarrow},
\label{m-spinor}
\end{align}
with the Pauli matrix $\tau_z$.

The mean-field Hamiltonian is diagonalized with a basis transformation
\begin{align}
W_{{\bm k}}^\dagger
h_{{\bm k}}
W_{{\bm k}}
=
\text{diag}\big(-E_{{\bm k}},E_{{\bm k}}\big),
\,\,\,\,
E_{{\bm k}}=\sqrt{\epsilon_{{\bm k}}^2+\Delta^2},
\end{align}
where 
\begin{align}
&W_{{\bm k}}
=\begin{pmatrix} 
\cos\theta_{{\bm k}} & \sin\theta_{{\bm k}}
\\
-\sin\theta_{{\bm k}}& \cos\theta_{{\bm k}}
\end{pmatrix},
\,\,\,\,
\tan\big(2\theta_{{\bm k}}\big)=\frac{\epsilon_{{\bm k}}}{\Delta}.
\label{gegeheje;e}
\end{align}
In the following, we use the abbreviations $\cos\theta_{{\bm k}}=c_{{\bm k}}$ and $\sin\theta_{{\bm k}}=s_{{\bm k}}$.
Without loss of generality, we choose the order parameter $\Delta$ to be positive,and therefore $\theta_{{\bm k}}\in(0,\pi/4)$, $s_{{\bm k}}>0$, and $c_{{\bm k}}=\sqrt{1-c_{{\bm k}}^2}$.  With this, the Hamiltonian  \eqref{ham:mf} becomes
\begin{align}
H_{mf}
&=\sum_{{\bm k},\sigma} 
\eta_{{\bm k},\sigma}^\dagger 
\text{diag}\big(-E_{{\bm k}},E_{{\bm k}}\big)
\eta_{{\bm k},\sigma} 
\\
&=\sum_{{\bm k},\sigma,\gamma=\pm} 
\gamma E_{{\bm k}}   b_{{\bm k},\sigma,\gamma}^\dagger   b_{{\bm k},\sigma,\gamma},
\end{align}
in terms of the new fermions
\begin{align}
\eta_{{\bm k},\sigma} 
&=
W_{{\bm k}}^\dagger
\Psi_{{\bm k},\sigma}
\equiv
\begin{pmatrix}
b_{{\bm k},\sigma,-}
\\
b_{{\bm k},\sigma,+}
\end{pmatrix}.
\label{gegenje223j3}
\end{align}
The order parameter Eq.~\eqref{m-spinor} is then
\begin{align}
m
&=
\frac{1}{2L}\sum_{{\bm k}}
\Big\langle
\eta_{{\bm k}\uparrow}^\dagger
W_{\bm k}^\dagger
\tau_z
W_{\bm k}
\eta_{{\bm k}\uparrow}
\Big\rangle
\\
&=
\frac{1}{2L}\sum_{{\bm k}}
(c_{{\bm k}}^2-s_{{\bm k}}^2)
\big(
n_{{\bm k}\uparrow,-}
-
n_{{\bm k}\uparrow,+}
),
\label{gweekee333}
\end{align}
where $n_{{\bm k}\sigma,\gamma}=\langle  b_{{\bm k}\sigma,\gamma}^\dagger b_{{\bm k}\sigma,\gamma} \rangle$. 
Using Eq.~\eqref{gegeheje;e} we obtain 
\begin{align}
c_{\bm k}^2-s_{\bm k}^2=
\cos(2\theta_{{\bm k}})=\frac{1}{\sqrt{\tan(2\theta_{{\bm k}})^2+1}}=\frac{\Delta}{\epsilon_{\bm k}}.
\end{align}
Furthermore, we assume that the occupation can be given by a general energy-dependent function $F(E)$, i.e., $n_{{\bm k},\gamma} = F(\gamma E_{{\bm k}})$. With the momentum summation \eqref{dos},  Eq.~\eqref{gweekee333} then gives the closed equation for the steady mean-field state,
\begin{align}
\label{fehege01119}
\frac{1}{U}
=
\int_0^\infty d\epsilon \,D_0(\epsilon)
\frac{F(-E)-F(E)}{E},\,\,\,\,\,E=\sqrt{\epsilon^2+\Delta^2}
\end{align}
which is the standard gap equation referenced in the main text.

\subsection{Solution of the gap equation for a state with two chemical potentials}

\begin{figure}[tbp]
	\includegraphics[width=0.5\textwidth]{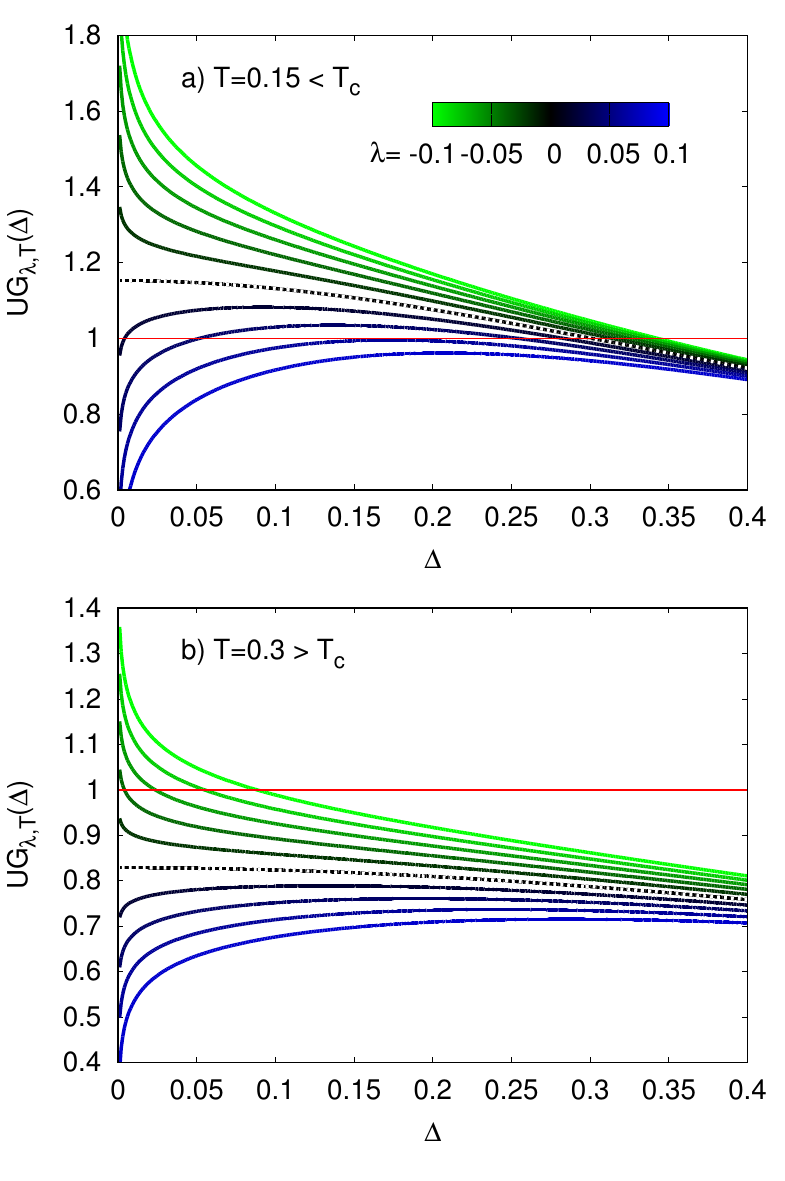}
	\caption{
		\label{fig:gdelta}
		The function $UG_{\lambda,T}(\Delta)$ for various $\lambda$ at $U=1.5$ and two temperatures $T>T_c$ and $T<T_c$. 
		The value $UG_{\lambda,T}(\Delta)=1$ corresponds to the physical gap. The dashed line corresponds to $\lambda=0$ (equilibrium); it has an intersection with $1$ for $T<T_c$, but not for $T>T_c$. For $\lambda<0$, $UG_{\lambda,T}(\Delta)$ has a logarithmic divergence at $\Delta=0$, so there is always a solution with a gap larger than the equilibrium gap at the same $T$. In contrast, for $\lambda<0$ there are two solutions with a reduced gap, or no solution.
	}
\end{figure}

As motivated in the main text, we solve the gap equation for a state with a different chemical potential in the upper and lower band, i.e., a distribution function 
\begin{align}
\label{gejeke}
F_{\lambda,T}(E) 
&= 
\begin{cases}
\frac{1}{e^{\beta(E-\lambda)}+1} & E>0
\\
\frac{1}{e^{\beta(E+\lambda)}+1} & E<0
\end{cases}.
\end{align}
With this, the gap equation \eqref{fehege01119} reads
\begin{align}
\frac{1}{U}
\stackrel{!}{=}
G_{\lambda,T}(\Delta)
\equiv
\int_0^\infty d\epsilon D_0(\epsilon) \frac{\tanh\Big(\frac{\sqrt{\epsilon^2+\Delta^2}-\lambda}{2T}\Big)}{\sqrt{\epsilon^2+\Delta^2}}.
\label{gdhjkeghjklw}
\end{align}
Other than in equilibrium, this equation can have multiple solutions.
For $\lambda=0$, Eq.~\eqref{gdhjkeghjklw} is the standard equilibrium gap equation, which has precisely one solution $\Delta_{eq}(T)>0$ for temperatures $T<T_c$ below a critical temperature, and no solution for $T>T_c$. By analyzing the behavior of the function $G_{\lambda,T}(\Delta)$ at $\Delta\to 0$ and $\Delta\to\infty$ (see Fig.~\ref{fig:gdelta}), one can see that for $\lambda<0$, there is always  one solution $\Delta(\lambda,T)>0$ with $\Delta(\lambda,T)>\Delta_{eq}(T)$ (even for $T>T_c$). For $\lambda>0$, there are either two solutions with $\Delta(\lambda,T)<\Delta_{eq}$, or no solution. In physical terms, the total occupation in the upper band can be obtained with a a positive $\lambda>0$ and a lower $T$, which concentrates the occupation at the bottom of the band, or a negative $\lambda<0$ and a larger $T$, which broadly distributes the occupation over the band and therefore is less harmful for the order parameter. In this work, we focus on the regime $\lambda<0$ (which describes the prethermal state, see main text), for which the phase diagram is shown in the main text. 

The total occupation in the upper band, and the total energy, are given by ($E=\sqrt{\epsilon^2+\Delta(\lambda,T)^2}$)
\begin{align}
N_{+}(\lambda,T) 
&= 4\int_0^{\infty} \!d\epsilon \,D_0(\epsilon) 
F_{\lambda,T}(E)
\label{mfnplus}
\\
\mathcal{E}(\lambda,T),
&=
\frac{\Delta^2}{U}-
4\int_0^{\infty} \!d\epsilon \, D_0(\epsilon)
E
\tanh\Big(\frac{E-\lambda}{2T}\Big),
\label{mfenergy}
\end{align}
where the factor $4$ is due to the spin summation and the normalization in Eq.~\eqref{dos}. The term $\frac{\Delta^2}{U}$ in the energy $\mathcal{E}$ is due the constant terms in the mean-field decoupling.

\section{Boltzmann equation}

\subsection{Scattering integral}

In the slow scattering approach, we assume that the time-evolving state is well described by a mean-field steady state, i.e., any ensemble of occupation number states, while these occupations slowly change due to scattering beyond the mean-field description. Fermi's golden rule for the scattering rate between two occupation number states $|\underline{n}\rangle=|\{n_{{\bm k},\gamma,\sigma}\}\rangle$ and  $|\underline{n}'\rangle=|\{n_{{\bm k},\gamma,\sigma}'\}\rangle$ in the mean-field basis gives the rate
\begin{align}
\Gamma_{\underline{n}\to\underline{n}'}
=
\frac{2\pi}{\hbar}
\big|\langle\underline{n}|H_{int}|\underline{n}'\rangle\big|^2 \delta(E_{\underline{n}}-E_{\underline{n}'}).
\end{align}
The interaction relates states $|\underline{n}\rangle$ and $|\underline{n}'\rangle$ which differ in the occupation of two orbitals with spin $\uparrow$ and two orbitals with spin $\downarrow$. In the following, it will be convenient to use a shorthand notation $j\equiv(\bm k_j,\gamma_j)$, $j=1,2,3,4$ for these orbitals, as well as  $\sum_j=\sum_{\bm k_j}\sum_{\gamma_j}$,  $E_j=\gamma_j\sqrt{\epsilon_{\bm k_j}^2+\Delta^2}$. Furthermore, refer to the scattering process  $(i,\uparrow)\to(j,\uparrow)$ and $(k,\downarrow)\to(l,\downarrow)$ as the scattering $(ij;kl)$. 

For the change of the occupation function in a given orbital $n_{1,\uparrow}=\langle b_{\bm k_1,\gamma_1,\uparrow}^\dagger b_{\bm k_1,\gamma_1,\uparrow} \rangle $ one therefore must sum over all initial and final states with scattering processes that change the occupation of $(1,\uparrow)$,
\begin{align}
\label{inout}
\frac{dn_{1,\uparrow}}{dt}
&=\Big(\frac{dn_{1,\uparrow}}{dt}\Big)_{in}-\Big(\frac{dn_{1,\uparrow}}{dt}\Big)_{out}
\\
\Big(\frac{dn_{1,\uparrow}}{dt}\Big)_{out}
&=
\frac{2\pi}{\hbar L^2}
\sum_{2,3,4}
|M_{1,2;3,4}|^2
F_{1,\uparrow}
\bar F_{2,\uparrow}
F_{3,\downarrow}
\bar F_{4,\downarrow}
\,\,\,\times
\nonumber\\
&\times\,\,\,\,
\delta_{{\bm k}_1+{\bm k}_3-{\bm k}_2-{\bm k}_4}
\delta (E_{1}+E_{3}-E_{2}-E_{4})
\nonumber\\
\Big(\frac{dn_{1,\uparrow}}{dt}\Big)_{in}
&=
\frac{2\pi}{\hbar L^2}
\sum_{2,3,4}
|M_{2,1;3,4}|^2
F_{2,\uparrow}
\bar F_{1,\uparrow}
F_{3,\downarrow}
\bar F_{4,\downarrow}
\,\,\,\times
\nonumber\\
&\times\,\,\,\,
\delta_{{\bm k}_2+{\bm k}_3-{\bm k}_1-{\bm k}_4}\delta (E_{2}+E_{3}-E_{1}-E_{4}).
\nonumber
\end{align}
Here the terms $(dn_{1,\uparrow}/dt)_{in}$ sum scattering processes $(21;34)$ into the state $(1,\uparrow)$, and 
$(dn_{1,\uparrow}/dt)_{out}$ sum scattering processes $(12;34)$ out of the state $(1,\uparrow)$. A scattering process $(ij;kl)$ is weighted with occupation factors for the initial orbitals $i$ and $k$ and $\bar F=(1-F)$ for the final orbitals $j$ and $l$ ($F_j\equiv F(E_j)$). Throughout this analysis, we will assume spin symmetry and spatial isotropy, i.e., the occupation depends only on energy. Finally, the matrix element $M_{ij;kl}$ for the process is obtained by representing  $H_{int}$ in the mean-field orbitals,
\begin{align}
H_{int}
=
\frac{1}{L}\sum_{ijkl}
M_{ij;kl}\,
\delta_{{\bm k}_j+{\bm k}_l-{\bm k}_i-{\bm k}_k}\,
b_{\bm k_j\gamma_j\uparrow}^\dagger
b_{\bm k_i\gamma_i\uparrow}
b_{\bm k_l\gamma_l\downarrow}^\dagger
b_{\bm k_k\gamma_k\downarrow}.
\label{hint-mf}
\end{align}
Explicit expressions for the matrix elements are given in Sec.~\ref{Uscatmf} below.

We then use this standard scattering theory to write down the change of the occupation density in the upper band,
\begin{align}
N_{+}\equiv\frac{1}{L}\sum_{\bm k_1,\sigma} n_{\bm k_1,+,\sigma} 
= \frac{2}{L}\sum_{\bm k_1} n_{\bm k_1,+,\uparrow}.
\end{align}
In the second equality we have taken  into account spin symmetry. Hence
\begin{align}
\frac{dN_{+}}{dt}
=
\frac{2}{L}\sum_{1} 
\delta_{\gamma_1,+} \Big[\Big(\frac{dn_{1,\uparrow}}{dt}\Big)_{in}-\Big(\frac{dn_{1,\uparrow}}{dt}\Big)_{out}\Big]
\end{align}
When inserting \eqref{inout} for $(dn_{1,\uparrow}/dt)_{in}$ in this expression, one can therefore exchange the summation variable $1$ and $2$, which gives
\begin{align}
\frac{dN_{+}}{dt}
&=
\frac{4\pi}{L^3\hbar}
\sum_{1,2,3,4}
(\delta_{\gamma_2,+}-\delta_{\gamma_1,+})
|M_{1,2;3,4}|^2
\delta_{{\bm k}_1+{\bm k}_3-{\bm k}_2-{\bm k}_4}
\,\,\,\times
\nonumber\\
&\times\,\,\,\,
F_{1}
\bar F_{2}
F_{3}
\bar F_{4}
\delta (E_{1}+E_{3}-E_{2}-E_{4}).
\nonumber
\end{align}
Using $\delta_{\gamma,+}=(\gamma+1)/2$, we have $(\delta_{\gamma_2,+}-\delta_{\gamma_1,+})=(\gamma_2-\gamma_1)/2$. Further symmetrizing the expression by exchanging $12$ and $34$, one can replace $(\delta_{\gamma_2,+}-\delta_{\gamma_1,+})=\Delta_{12;34}/2$, with  
\begin{align}
\Delta_{12;34} = \frac{\gamma_2-\gamma_1+\gamma_4-\gamma_3}{2}.
\end{align}
The factor $\Delta_{12;34}$ intuitively distinguishes between different scattering processes, where $(\gamma_1,\gamma_2,\gamma_3,\gamma_4)$ takes the following combinations:
\begin{itemize}
	\item[(1)]
	Intra-band scattering: $(++++)$, $(----)$;
	\item[(2)]
	Scattering of particles in upper band on particles in lower band: $(++--)$, $(--++)$ ;
	\item[(3)]
	Exchange of particles between bands: $(+--+)$, $(-++-)$;
	\item[(4)] Auger processes: 
	$(-+++)$, $(+-++)$, $(++-+)$, $(+++-)$, $(+---)$, $(-+--)$, $(--+-)$, $(---+)$.
\end{itemize}
Processes (1) would lead to separate thermalization of electrons in the upper and lower band, processes (2) and (3) would equilibrate the temperatures of the upper and lower band. Obviously,  $\Delta_{12;34}=0$ for processes (1), (2), and (3), so that only Auger processes with $\Delta_{12;34}=\pm1$ contribute to a change in the upper band population. 

Finally, to derive the final expression of the main text, we further neglect the momentum conservation, which is consistent with the infinite dimensional limit of DMFT,
\begin{align}
\label{ratefinal}
\frac{dN_+}{dt}
&=
\frac{2\pi}{\hbar L^4}
\sum_{1,2,3,4}
\Delta_{12;34}
|M_{1,2;3,4}|^2 \,\,\,\,\times
\nonumber\\
&\times \,\,\,\,F_{1}  \bar F_{2} F_{3}  \bar F_{4}
\delta (E_{1}+E_{3}-E_{2}-E_{4}).
\end{align}

\subsection{Representation of the interaction in the mean-field basis}
\label{Uscatmf}

To derive the matrix elements in Eq.~\eqref{hint-mf}, we use the inverse of the  transformation defined by Eqs.~\eqref{gegenje223j3} and \eqref{gegeheje;e}, which can be written as
\begin{align}
c_{{\bm k},\alpha,\sigma}
=
\sum_{\gamma=\pm}
u_{{\bm k},\sigma}^{\alpha,\gamma}  \, b_{{\bm k},\sigma,\gamma}
\label{ga;ebehelde}
\end{align}
with 
\begin{align}
\begin{pmatrix}
u_{{\bm k},\uparrow}^{A,-}
& 
u_{{\bm k},\uparrow}^{A,+}
\\
u_{{\bm k},\uparrow}^{B,-}
&
u_{{\bm k},\uparrow}^{B,+}
\end{pmatrix}
&=
\begin{pmatrix}
c_{{\bm k}} & s_{{\bm k}}
\\
-s_{{\bm k}}& c_{{\bm k}}
\end{pmatrix}
\\
\begin{pmatrix}
u_{{\bm k},\downarrow}^{A,-}
& 
u_{{\bm k},\downarrow}^{A,+}
\\
u_{{\bm k},\downarrow}^{B,-}
&
u_{{\bm k},\downarrow}^{B,+}
\end{pmatrix}
&=
\begin{pmatrix}
-s_{{\bm k}} & c_{{\bm k}}
\\
c_{{\bm k}}& s_{{\bm k}}
\end{pmatrix}.
\end{align}
Transforming Eq.~\eqref{ga;ebehelde} to real space,  for a site $j$ on sublattice $\alpha$, 
\begin{align}
c_{j,\sigma} 
&= 
\frac{1}{\sqrt{L}}\sum_{{\bm k}} e^{i{\bm k}{\bm r}_j} \sum_{\gamma=\pm}  u_{{\bm k},\sigma}^{\alpha,\gamma}\,b_{{\bm k},\sigma,\gamma},
\end{align}
the interaction Hamiltonian $H_{int}=\sum_{j}c_{j\uparrow}^\dagger c_{j\uparrow} c_{j\downarrow}^\dagger c_{j\downarrow}$ is written in the form \eqref{hint-mf}, with 
\begin{align}
&M_{1,2;3,4}
=
U\sum_{\alpha}
u_{{\bm k}_1,\uparrow}^{\alpha,\gamma_1}
u_{{\bm k}_2,\uparrow}^{\alpha,\gamma_2}
u_{{\bm k}_3,\downarrow}^{\alpha,\gamma_3}
u_{{\bm k}_4,\downarrow}^{\alpha,\gamma_4},
\end{align}
where we have used the fact the all $u_{{\bm k},\sigma}^{\alpha,\gamma}$ are real-valued. We can now write down the matrix elements for all possible combinations of $\gamma$-indices, with the obvious notation
$M_{1,2,3,4}^{\gamma_1,\gamma_2,\gamma_3,\gamma_4}=M_{({\bm k}_1,\gamma_1)({\bm k}_2,\gamma_2);({\bm k}_3,\gamma_3)({\bm k}_4,\gamma_4)}$, $s_j=s_{{\bm k}_j}$, $c_j=c_{{\bm k}_j}$:
\begin{align}
\label{gstrrf01}
&M_{1,2,3,4}^{++++}=M_{1,2,3,4}^{----}=U(s_{1}s_{2}c_{3}c_{4}+c_{1}c_{2}s_{3}s_{4}),\\
\label{gstrrf02}
&M_{1,2,3,4}^{--++}=M_{1,2,3,4}^{++--}=U(c_{1}c_{2}c_{3}c_{4}+s_{1}s_{2}s_{3}s_{4}),\\
\label{gstrrf03}
&M_{1,2,3,4}^{+-+-}=M_{1,2,3,4}^{-+-+}=-U(s_{1}c_{2}c_{3}s_{4}+c_{1}s_{2}s_{3}c_{4}),\\
\label{gstrrf04}
&M_{1,2,3,4}^{+--+}=M_{1,2,3,4}^{-++-}=-U(s_{1}c_{2}s_{3}c_{4}+c_{1}s_{2}c_{3}s_{4}),\\
\label{gstrrf05}
&M_{1,2,3,4}^{-+++}=-M_{1,2,3,4}^{+---}=U(c_{1}s_{2}c_{3}c_{4}-s_{1}c_{2}s_{3}s_{4}),\\
\label{gstrrf06}
&M_{1,2,3,4}^{+-++}=-M_{1,2,3,4}^{-+--}=U(s_{1}c_{2}c_{3}c_{4}-c_{1}s_{2}s_{3}s_{4}),\\
\label{gstrrf07}
&M_{1,2,3,4}^{++-+}=-M_{1,2,3,4}^{--+-}=U(-s_{1}s_{2}s_{3}c_{4}+c_{1}c_{2}c_{3}s_{4}),\\
\label{gstrrf08}
&M_{1,2,3,4}^{+++-}=-M_{1,2,3,4}^{---+}=U(-s_{1}s_{2}c_{3}s_{4}+c_{1}c_{2}s_{3}c_{4}).
\end{align}
These relations will be used in the implementation below.

In passing we note that following Eq.~\eqref{gegeheje;e}, the factors $s_j$ tend to zero for states  at the bottom of the valence band and top of the conduction band, where $\epsilon_{\bm k}=0$. Hence for distributions with electrons and holes centered at the conduction band minimum and valence band top, respectively, the matrix element for process \eqref{gstrrf02}, which describes scattering of a hole on an electron and vice versa, is of order one, while the matrix elements for Auger processes \eqref{gstrrf05}-\eqref{gstrrf08} are of order $s_j$ and therefore much smaller. The physical reason is that states with the same spin at the valence band top and conduction band minimum have their weight on different sublattices, and are therefore not coupled through a local interaction. This fact supports the separation of timescales between the individual thermalization of electrons and holes, mediated by processes \eqref{gstrrf02}, and the redistribution of weight between the bands, mediated by Auger processes, which is observed in the full DMFT results presented in the main text. A more important reason, however, independent of the detailed form of the interaction, should be phase-space restrictions due to the gap.

\subsection{Slow variable dynamics}

We now calculate a dynamics under the assumption that the intra-band scattering is fast enough such that at each instance of time, the system is in a well-defined non-equilibrium state with a distribution function \eqref{gejeke} characterized by parameters $\lambda(t)$ and $T(t)$. The dynamics of these slow variables is therefore determined by the rate of change of $N_+$ and the conservation of the energy \eqref{mfenergy}
\begin{align}
\frac{d \mathcal{E}(\lambda,T)}{dt}&=0,
\label{gehejjjj01}
\\
\frac{d N_{+}(\lambda,T)}{dt} &= \Gamma(\lambda,T).
\label{gehejjjj02}
\end{align}
Here $\Gamma(\lambda,T)$ is the rate \eqref{ratefinal} evaluated with the distribution function \eqref{gejeke}. 
Moreover, the gap  $\Delta(t)$ at each instance of time is determined by the gap equation \eqref{gdhjkeghjklw}. 
This set of equations can be solved by various equivalent means. Here we proceed as follows: To deal with the constraint \eqref{gdhjkeghjklw}, we can also view $G$, $N_{+}$, and $\mathcal{E}$ 
as function of $\lambda$, $T$, and $\Delta$. The total derivatives \eqref{gehejjjj01} and \eqref{gehejjjj02}, together with $dG/dt=0$ can be written in the form 
\begin{align}
\begin{pmatrix}
G_{\lambda}  & G_{T} & G_\Delta
\\
\mathcal{E}_{\lambda}  & \mathcal{E}_{T} & \mathcal{E}_{\Delta}
\\
N_{\lambda}  & N_{T} & N_\Delta
\end{pmatrix} 
\begin{pmatrix}
\dot \lambda
\\
\dot T
\\
\dot \Delta
\end{pmatrix}
=
\begin{pmatrix}
0
\\
0
\\
\Gamma
\end{pmatrix},
\label{gehejeeehe77e7e}
\end{align}
with the short notation $X_y=\partial X/\partial y$. The various derivatives can be obtained explicitly in the form of integral expressions which are then numerically evaluated. For example, using Eq.~\eqref{mfnplus}
\begin{align}
\label{der01}
N_\Delta
&=4\int_0^{\infty} d\epsilon \,D_0(\epsilon) f'\Big( \frac{E-\lambda}{2T}\Big)\frac{\Delta}{2T\sqrt{\epsilon^2+\Delta^2}}.
\\
\label{der02}
N_\lambda
&=-4\int_0^{\infty} d\epsilon \,D_0(\epsilon) f'\Big( \frac{E-\lambda}{2T}\Big)\frac{1}{2T}.
\\
\label{der03}
N_T
&=-4\int_0^{\infty} d\epsilon \,D_0(\epsilon) f'\Big( \frac{E-\lambda}{2T}\Big)\frac{E-\lambda}{2T^2}.
\end{align}
We then eliminate $dt$ from the first two equations in \eqref{gehejeeehe77e7e},
\begin{align}
\begin{pmatrix}
G_{T} & G_\Delta
\\
\mathcal{E}_{T} & \mathcal{E}_{\Delta}
\end{pmatrix} 
\begin{pmatrix}
dT/d\lambda
\\
d\Delta/d \lambda
\end{pmatrix}
=
-
\begin{pmatrix}
G_{\lambda}  
\\
\mathcal{E}_{\lambda} 
\end{pmatrix}.
\label{gehejeeehe77e7e737}
\end{align}
Solving this equation gives the constant energy curve $T(\lambda)$ and the gap along the constant energy curve. Once Eq.~\eqref{gehejeeehe77e7e737} is solved, one can directly integrate the third equation in \eqref{gehejeeehe77e7e} to get the time along the trajectory,
\begin{align}
\label{egejekk2}
dt 
=
d\lambda
\,
\frac{N_\lambda + N_T \frac{dT}{d\lambda} + N_\Delta \frac{d\Delta}{d\lambda}}{\Gamma}.
\end{align}
In the time-evolution, $\lambda$ will evolve monotonously from a nonzero value to $0$ (equilibrium). To avoid a singular behavior of the derivative $dt/d\lambda$ related to a slow-down as $\lambda$ approaches $0$, one can use a parametrization 
\begin{align}
\Delta = \Delta(0) e^{-s}
\\
\lambda = \lambda(0) e^{-y},
\end{align}
so that \eqref{gehejeeehe77e7e} and and \eqref{gehejeeehe77e7e737} become
\begin{align}
\begin{pmatrix}
-\lambda G_{\lambda}  & G_{T} & -\Delta G_\Delta
\\
-\lambda \mathcal{E}_{\lambda}  & \mathcal{E}_{T} & -\Delta \mathcal{E}_{\Delta}
\\
-\lambda N_{\lambda}  & N_{T} & -\Delta N_\Delta
\end{pmatrix} 
\begin{pmatrix}
\dot y
\\
\dot T
\\
\dot s
\end{pmatrix}
=
\begin{pmatrix}
0
\\
0
\\
\Gamma
\end{pmatrix},
\\
\begin{pmatrix}
G_{T} & -\Delta G_\Delta
\\
\mathcal{E}_{T} & -\Delta\mathcal{E}_{\Delta}
\end{pmatrix} 
\begin{pmatrix}
dT/dy
\\
ds/d y
\end{pmatrix}
=
\begin{pmatrix}
\lambda G_{\lambda}  
\\
\lambda \mathcal{E}_{\lambda} 
\end{pmatrix}.
\end{align}

\subsection{Numerical aspects}
\label{secLNumerical}

To evaluate the rate \eqref{ratefinal}, we replace $\sum_j \to 2L\int_0^\infty d\epsilon_j D_0(\epsilon_j) \sum_{\gamma_j}$ [c.f.~Eq.~\eqref{dos}]. For an easy implementation of the $\delta$-function it is convenient to consider the integration variable $\bar E_j=\sqrt{\epsilon_j^2+\Delta^2}$. Energy integrals become
\begin{align}
&\int_0^4 d\epsilon D_0(\epsilon) g(E) = \int_{\Delta}^{E_{max}} dE \,\tilde D(E)
g(E) 
\end{align}
with $E_{max}=\sqrt{4+\Delta^2}$, and the transformed density of states
\begin{align}
\label{dostr}
\tilde D(E)=
\frac{D(E(\epsilon))}{|dE/d\epsilon|} =
\frac{1}{2\pi}
\frac{\sqrt{E_{max}^2-E^2}}{\sqrt{E^2-\Delta^2}/E}.
\end{align}
The rate \eqref{ratefinal} is then given by
\begin{align}
&\frac{dN_+}{dt}
=
\frac{2\pi}{\hbar}
\int_{\Delta}^{E_{max}}\!\!\!\!
dE_1 \tilde D(E_1) \cdots dE_4\tilde D(E_4)
\nonumber\\
&\times \,\,\,\,\,
\sum_{\gamma_1,\gamma_2,\gamma_3,\gamma_4}\!\!\!\!
\Delta_{12;34} F(\gamma_1E_1)  \bar F(\gamma_2E_2) F(\gamma_3E_3)  \bar F(\gamma_4E_4)
\nonumber\\
&\times \,\,\,\,\,
|M_{1,2;3,4}|^2
\delta (\gamma_1E_{1}-\gamma_2E_{2}+\gamma_3E_{3}-\gamma_4E_{4}).
\end{align}
Here $M_{1,2;3,4}$ is obtained from Eqs.~\eqref{gstrrf01}-\eqref{gstrrf08}, noting that Eq.~\eqref{gegeheje;e} implies
\begin{align}
c^2
=
\frac{1}{2}\Big[1+\sqrt{\frac{\Delta^2}{E^2}}\Big],
\,\,\,
s^2=\frac{1}{2}\Big[1-\sqrt{\frac{\Delta^2}{E^2}}\Big].
\label{kjeclnws;}
\end{align}
If we perform the integral over $E_1$ first, the other integrals are simply restricted to 
\begin{align}
\label{fegege}
E_{1}=\gamma_1(\gamma_2E_{2}-\gamma_3E_{3}+\gamma_4E_{4}).
\end{align}
Moreover, we use the parametrization $E=\Delta+ W\sin(x)^2$, $x\in(0,\pi/2)$, $W=(E_{max}-\Delta)$ in all $E$ integrals to lift the root-divergencies in the transformed density of states \eqref{dostr}, also in the evaluation of the derivatives such as \eqref{der01}-\eqref{der03}.

\section{DMFT and memory truncation}
\label{sec:DMFT}

In this section, we present the DMFT equations for the solution of model \eqref{ham:hubb}. The basic equations are  analogous to Ref.~\cite{Tsuji2013}. They are reproduced here in order to outline the memory truncation scheme employed in the present work to reach long times. 

Nonequilibrium DMFT \cite{Aoki2014} is a reformulation of DMFT on the Keldsyh contour $\mathcal{C}$. We aim to compute the contour-ordered Green's function 
\begin{align}
\hat G_{{\bm k},\sigma}(t,t') = -i \langle T_\mathcal{C} \psi_{{\bm k},\sigma}(t) \psi_{{\bm k},\sigma}^\dagger(t')\rangle.
\end{align}
(For an introduction to the Keldysh formalism and the notation, see, e.g., Ref.~\onlinecite{Aoki2014}.) The Green's functions has a $2\times2$ matrix structure in orbital space, introduced through the spinors \eqref{spinors}. We use the hat supercript to indicate such $2\times2$ matrices. Within the local approximation of DMFT, the Green's function depends on ${\bm k}$ only via the band energy $\epsilon_{{\bm k}}$, and we use the notation $\hat G_{\epsilon,\sigma}\equiv \hat G_{{\bm k},\sigma}$ for $\epsilon_{{\bm k}}=\epsilon$. Local Green's functions are obtained by momentum summation \eqref{dos}, or 
\begin{align}
\label{gloc}
\hat G_{\sigma}(t,t')
=
\int_0^\infty \!\!d\epsilon \,2D_0(\epsilon) \hat G_{\epsilon,\sigma}(t,t'),
\end{align}
and the diagonal components of this Green's function constitute the local Green's functions $G_{\alpha,\sigma}$ on sublattice $\alpha=A,B$. With the spinor \eqref{spinors}, the $00$ and $11$ component of $\hat G_{\uparrow}$ are $G_{A\uparrow}$ and $G_{B\uparrow}$, respectively, while the $00$ and $11$ component of $\hat G_{\downarrow}$ are $G_{B\downarrow}$ and $G_{A\downarrow}$. The expectation value of the order parameter \eqref{m-spinor} is thus given by
$m(t)=\text{tr}\big[\hat \tau_z \hat G_{\sigma}^<(t,t)\big]/(2i)$.

This Green's function satisfies the Dyson equation
\begin{align}
[i\partial_t +\mu 
&- \hat h_{\epsilon}(t)]\hat G_{\epsilon,\sigma}(t,t')
\nonumber\\
&-\int_{\mathcal{C}}\! d\bar t\, \hat\Sigma_\sigma(t,\bar t) \hat G_{\epsilon,\sigma}(\bar t, t') =\delta(t,t')
\label{dyson}.
\end{align}
Here the Hartree self-energy is incorporated into $h_{\epsilon}$ as in \eqref{ham:mf}, such that $\hat h_\epsilon(t)= -Um(t)\hat \tau_z + \epsilon \hat\tau_x$. The correlation self energy $\hat\Sigma_\sigma$ is diagonal on the lattice (momentum independent in the MBZ), $\hat \Sigma_{\uparrow} = \text{diag}(\Sigma_{A\uparrow},\Sigma_{B\uparrow})$ and 
$\hat \Sigma_{\downarrow} = \text{diag}(\Sigma_{B\downarrow},\Sigma_{A\downarrow})$, where the local self-energies $\Sigma_{\alpha,\sigma}$ are obtained from the auxiliary DMFT impurity model. Since we are interested in the dynamics of the weak-coupling Slater antiferromagnet, we solve the impurity model using second-order perturbation theory,
\begin{align}
\label{sigmaloc}
\Sigma_{\alpha,\sigma}=U(t)U(t') G_{\alpha,\sigma}(t,t') G_{\alpha,\bar\sigma}(t',t) G_{\alpha,\bar\sigma}(t,t'),
\end{align}
where $\bar\sigma\!=\,\uparrow$ when $\sigma\!=\,\downarrow$ and vice versa. With this, Eqs.~\eqref{gloc}-\eqref{sigmaloc} form a closed set self-consistent equations.

\subsection{Memory truncation}
\begin{figure}[tbp]
	\includegraphics[width=0.5\textwidth]{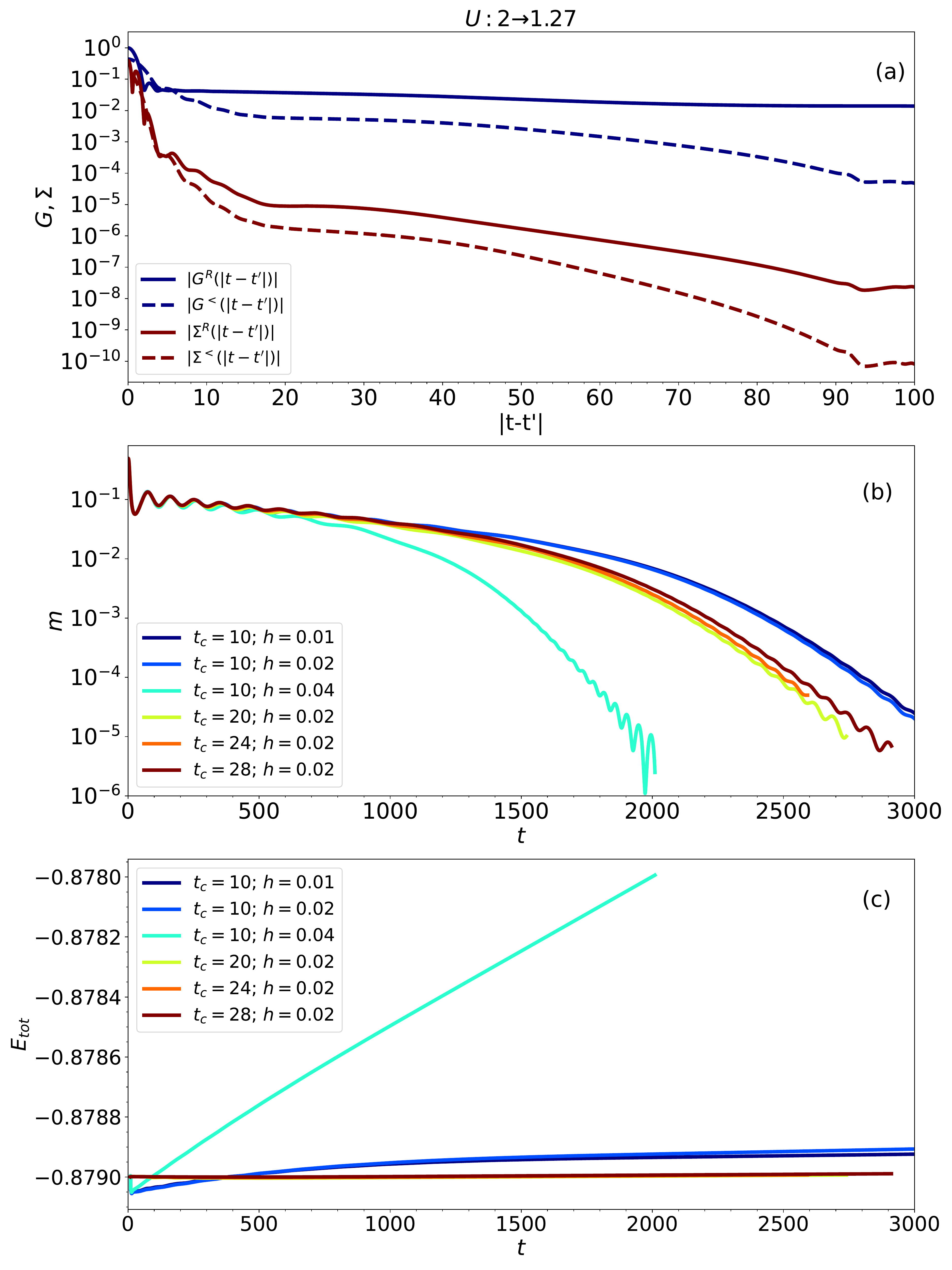}
	\caption{
		\label{fig:truncation}
		a) Self-energy and local Green's function  as a function of the relative time $|t-t'|$ for the quench $U:2 \to 1.27$. b) and c): Time evolution of $m(t)$ and $E_{tot}(t)$ for different values of cutoff time $t_c$ and time grid $h$.  
	}
\end{figure}
The main numerical challenge in long-time simulations are the integrals in \eqref{dyson}. However, as the correlation self-energy decays quickly as a function of relative time (Fig.~\ref{fig:truncation}a), we can set $\Sigma(t,t')=0$ when $|t-t'|>t_c$, with some cutoff time $t_c$.  With this, the numerical effort for the propagation of Eq.~\eqref{dyson} over a time-interval $t_{max}$ scales like $t_{max}t_c^2$, instead of $t_{max}^3$, and the required storage scales like $t_c^2$, instead of $t_{max}^2$. While the idea of this truncation is straightforward \cite{Schueler2018}, technicalities of the implementation, as well as a more detailed analysis how well it works in various physical systems will be presented  elsewhere. For the present analysis, the cutoff $t_c$ is varied at the end in order to confirm that the results are converged on all times. 

The convergence is analyzed exemplarily for the quench in $U$ from 2 to 1.27: In Fig.~\ref{fig:truncation}b we show the order parameter $m(t)$ and the total energy $E_{tot}(t)$, respectively, for different combinations of the cutoff $t_c$ and the time step $h$ used in the discretization of the convolution integrals and the derivatives in Eq.~\eqref{dyson}. Because the magnitude of the order parameter is already small in the long-time regime ($m\lesssim 10^{-3}$ during the accelerated gap collapse), the numerical data must have a high accuracy to resolve the dynamics. To reach sufficient accuracy, we rely on the time-propagation implemented within the NESSi simulation package \cite{Schueler2020}, for which  the numerical error scales better than $\mathcal{O}(h^5)$ with the timestep $h$. As a consequence, the data quickly converge as the timestep falls below some threshold (compare the curves  for $h=0.04,0.02,0.01$ at $t_c=10$ in Figs.~\ref{fig:truncation}b and ~\ref{fig:truncation}c). Once the time grid has been fixed ($h=0.02$ is considered sufficiently accurate for our purposes), $t_c$ is increased in order to make sure that the results have reached convergence. In Fig.~\ref{fig:truncation}b one can see that passing from $t_c=10$ to $t_c=20$ makes a difference in the time evolution of $m(t)$, whilst further increasing $t_c$ above $t_c=20$ does not significantly affect neither the total energy conservation nor the dynamics of $m(t)$. Another measure for the numerical accuracy of the results is the conservation of the total energy after the quench (see Fig.~\ref{fig:truncation}c). For the parameters that have been chosen in all the simulations ($t_c=20$ and $h=0.02$), the relative variation of the total energy after the quench is of the order of $10^{-6}$ for all  quenches and all  times, an order of magnitude that makes us confident that also the accelerated collapse of $m(t)$ is not a numerical artefact due to an artificial ``heating'' of the system.

The convergence of the results around a cutoff $t_c\approx 20$ is consistent with the decay of the self-energy shown in  Fig.~\ref{fig:truncation}a: for $|t-t'|=10$, $|\Sigma^R(t-t')|$ is around $10^{-4}$; for $|t-t'|=20$, $|\Sigma^R(t-t')|$ is around $10^{-5}$. From the behavior of $|\Sigma^R(t-t')|$ at larger times, one would expect a further substantial increase of the accuracy if $t_c$ is increased beyond $t=40$ (a factor of two).  This would increase both the computation time and the required memory by a factor of four. 

We remark that for a semielliptic density of states, the exact  DMFT equations can be written in a closed form equation for the local Green's functions,
\begin{align}
&[i\partial_t +\mu 
- \xi_\alpha\sigma m U]\hat G_{\alpha,\sigma}(t,t')
\nonumber\\
&-\int_{\mathcal{C}}\! d\bar t\, [\Delta_{\alpha,\sigma}(t,\bar t) + \Sigma_{\alpha,\sigma}(t,\bar t)] \hat G_{\alpha,\sigma}(\bar t, t') =\delta(t,t'),
\label{dyson-closed}
\end{align}
with a hybridization function $\Delta_{A,\sigma}= t_h^2 G_{B,\sigma}$, $\Delta_{B,\sigma}= t_h^2 G_{A,\sigma}$. However, while the self-energy decays quickly with relative time, the Green's function does not (Fig.~\ref{fig:truncation}a): In the ordered phase, the spectral function has a $1/\sqrt{\epsilon}$ van Hove singularity at the gap edges, which translates to a slow decay. Hence one would have to  use much larger cutoff times $t_c$ when truncating the memory integrals in Eq.~\eqref{dyson-closed} instead of Eq.~\eqref{dyson}. For the results shown here the summation \eqref{gloc} was performed using $N_\epsilon=640$ values of $\epsilon$, and the Dyson equation \eqref{dyson} was solved for each $\epsilon$ with a truncation of the $\Sigma$ integrals.

%

\end{document}